\documentstyle[12pt]{article}
\def\be{\begin{equation}}
\def\ee{\end{equation}}
\def\bea{\begin{eqnarray}}
\def\eea{\end{eqnarray}}

\begin{document}

\begin{center}
{\Large{\bf Noncommutativity of the Moving $D2$-brane Worldvolume}}

\vskip .5cm
{\large Davoud Kamani}
\vskip .1cm
 {\it Institute for Studies in Theoretical Physics and
Mathematics (IPM)
\\  P.O.Box: 19395-5531, Tehran, Iran}\\
{\sl e-mail: kamani@theory.ipm.ac.ir}
\\
\end{center}

\begin{abstract}

In this paper we study the noncommutativity of 
a moving membrane with background fields. The open string variables are 
analyzed. Some scaling limits are studied. The equivalence of the magnetic
and electric noncommutativities is investigated.
The conditions for equivalence of
noncommutativity of the $T$-dual theory in the rest frame and
noncommutativity of the original theory in the moving frame are obtained. 
                                                                       
\end{abstract}

\newpage
\section{Introduction}
Over the past years there have been attempts to explain 
noncommutativity on D-brane worldvolume through the study of
open strings in the presence of background fields
\cite{1,2}. From the DBI action
it is known that on a D-brane electric field cannot be stronger
than a critical electric field, while the same is not true  
for magnetic fields. Also it is known that Lorentz boosts act on 
electromagnetic backgrounds. This affects the noncommutativity 
parameter and the effective open string metric.
According to these facts, some properties of D-branes with
background fields such as decoupling limits and light-like
noncommutativity have been studied \cite{3}.

Previously we have studied the noncommutativity of a moving
$Dp$-brane, with the motion along itself \cite{4}. Now we
study the noncommutativity of a moving membrane
with electric and magnetic background fields. The motion is parallel or
perpendicular to the membrane. For each case, the effective open string
variables will be analyzed. We shall observe that for an appropriate
magnetic field the open string metric is frame independent.
In a special frame the open string metric is proportional to the
closed string metric.
For both electric and magnetic cases we find decoupling limits,
which lead to the definite noncommutative theories.
There are frames for the electric and magnetic membranes such that
their noncommutativities are proportional to each other.

The effects of $T$-duality on the effective metric and 
noncommutativity parameter enable us to obtain equivalent
noncommutativity structures. That is, we find
speeds and background fields for the membrane such that the 
noncommutativity of the $T$-dual theory becomes equivalent to the 
noncommutativity of the original theory in the moving frame.

The analysis of Ref.\cite{2} leads to the definitions of
the open string metric $G$, the noncommutativity parameter $\Theta$ and 
the effective open string coupling constant $G_s$,
\bea
&~& G^{\mu \nu} = \bigg{(}(g+2 \pi \alpha' B)^{-1}g (g-2 \pi
\alpha' B)^{-1} \bigg{)}^{\mu \nu}\;, 
\nonumber\\
&~& G_{\mu \nu} = \bigg{(}(g-2 \pi \alpha' B)g^{-1} (g+2 \pi
\alpha' B) \bigg{)}_{\mu \nu}\;, 
\nonumber\\
&~& \Theta^{\mu \nu} = -(2 \pi \alpha')^2 \bigg{(}(g+2 \pi
\alpha' B)^{-1} B (g-2 \pi
\alpha' B)^{-1} \bigg{)}^{\mu \nu}\;,
\nonumber\\
&~& G_s = g_s \bigg{(} \frac{\det G}{\det
(g+2\pi \alpha'B)}\bigg{)}^{\frac{1}{2}} \;,
\eea
where $g_{\mu \nu}, B_{\mu \nu}$ and $g_s$ are closed string variables.

Note that the effective open string coupling $G_s$ does not
change under the Lorentz boosts. Because $g_s$ is the exponential of the 
scalar field dilation, and the ratio of two determinants 
also is invariant.

In general a $D2$-brane parallel to the $X^1 X^2$-plane has the 
NS-NS background $B$-field as in the following 
\bea
B_{\mu \nu} = \left( \begin{array}{ccc}
0 & E & E'\\
-E & 0 & b\\
-E' & -b & 0
\end{array} \right)\;,
\eea
where $\mu , \nu \in \{0,1,2 \}$. 
We shall discuss pure magnetic and pure electric cases.
Let the closed string metric of the membrane worldvolume be
\bea
g_{\mu \nu} = \left( \begin{array}{ccc}
-g_0 & 0 & 0 \\
0 & g_1 & g'\\
0 & g' & g_2
\end{array} \right).
\eea

The paper is organized as follows. In section 2, we study the behavior 
of effective variables of open string
in terms  of the background magnetic field and the 
speed of the membrane. In section 3, the same will be done in the
presence of the electric field. In addition, some scaling limits and
also equivalence of two noncommutativities will be obtained.
In section 4, we study the $T$-duality of the theory and
conditions for equivalence of boosted and T-dual noncommutativities.
\section{Magnetic field noncommutativity}

For the pure magnetic field (i.e., $E=E'=0$),
the noncommutativity matrix is
\bea
\Theta^{\mu \nu} = \theta \left( \begin{array}{ccc}
0 & 0 & 0 \\
0 & 0 & 1 \\
0 & -1 & 0
\end{array} \right)\;,
\eea
where the parameter $\theta$ (i.e., the strength of the noncommutativity)
is defined by
\bea
&~& \theta = - \frac{(2\pi \alpha')^2 b}{g+(2\pi \alpha')^2 b^2}\;,
\nonumber\\
&~& g \equiv  g_1g_2 -g'^2 \;.
\eea
This relation implies that the different magnetic fields
$b_\pm = -\frac{1}{2\theta} \pm \frac{1}{2} \sqrt{\frac{1}{\theta^2}
-\frac{g}{\pi^2 \alpha'^2}}$ produce the same noncommutativity on the 
membrane. Since $g$ is positive, the function $\theta (b)$ 
has the maximum $\theta_0$ (the minimum $-\theta_0$)
at $b=-b_0$, $(b=b_0)$ where
\bea
b_0 = \frac{\sqrt{g}}{2\pi \alpha'}\;\;\;\;,\;\;\;
\theta_0 = \frac{\pi \alpha'}{\sqrt{g}}\;.
\eea
Therefore, to obtain a strong noncommutativity we should
adopt $\pm b_0$ magnetic fields.

Now we proceed to study the expressions for the various geometrical
quantities in different frames, appropriate to different states of motion
of the brane.
\subsection{Motion along the $X^1$-direction}

Consider the Lorentz transformations on the coordinates $X^0$ and $X^1$,
\bea
&~& X'^1 = \gamma (X^1 - v \sqrt{g_0}X^0) \;,
\nonumber\\
&~& X'^0 = \gamma (X^0 - \frac{v}{\sqrt{g_0}}X^1) \;,
\eea
where $\gamma = 1/\sqrt{1-v^2}$. The effect of these
transformations on the matrix (4) is
\bea
\Theta'^{\mu \nu} = \gamma \theta \left( \begin{array}{ccc}
0 & 0 & -\frac{v}{\sqrt{g_0}} \\
0 & 0 & 1 \\
\frac{v}{\sqrt{g_0}} & -1 & 0
\end{array} \right)\;.
\eea
On the other hand, in the moving frame the noncommutativity parameter
has non-zero time-like element.
Since for a brane with electric field also there are time-like elements
(e.g., see the equation (19)), this implies that the motion
along the brane directions is
equivalent to an appropriate electric field.

The transformation of the open string metric is
\bea
G'_{\mu \nu} = \left( \begin{array}{ccc}
-\gamma^2 g_0(1-g_1 av^2) & -\gamma^2 v \sqrt{g_0}(1-g_1 a) & 
\gamma v \sqrt{g_0}g'a \\
-\gamma^2 v \sqrt{g_0}(1-g_1 a) & \gamma^2 (g_1 a-v^2) & \gamma g'a\\
\gamma v \sqrt{g_0}g'a & \gamma g'a & g_2 a
\end{array} \right)\;,
\eea
where the parameter $a$ is defined by
\bea
a=1+\frac{(2\pi \alpha' b)^2}{g}\;.
\eea
Since $g$ is positive we have $a \geq 1$.
For $v=0$, this metric reduces to the open string metric for the static
membrane with magnetic field.

Let the off-diagonal element $g'$ vanish. In addition, consider the
following relation between the magnetic field and the elements of
the closed string metric
\bea
b = \pm \frac{1}{2\pi \alpha'} \sqrt{g_2 (1-g_1)}\;.
\eea
These give the diagonal open string metric 
\bea
G'_{\mu\nu} = {\rm diag} (-g_0 \;,\; 1 \;,\; g_2/g_1)\;,
\eea
which is independent of the speed of the membrane. Note that the membrane
speed has not been hidden in the relation (11).

Under the above conditions we have
$\theta = \pm 2\pi \alpha' \sqrt{\frac{1-g_1}{g_2}}$. Therefore,
for $v \rightarrow 1$, $g_1 \rightarrow 1$ (i.e., $b \rightarrow 0$ )
and finite $g_0$ and $g_2$ we can introduce the scaling limit
\bea
\gamma \sqrt{1-g_1} = \sigma\;,
\eea
where $\sigma$ is a finite constant. This implies that
\bea
&~& \Theta'^{\mu \nu} = \pm 2\pi \alpha' \frac{\sigma}{\sqrt{g_2}}
\left( \begin{array}{ccc}
0 & 0 & -\frac{1}{\sqrt{g_0}} \\
0 & 0 & 1 \\
\frac{1}{\sqrt{g_0}} & -1 & 0
\end{array} \right)\;,
\nonumber\\
&~& G'_{\mu\nu} = {\rm diag} (-g_0 \;,\; 1 \;,\; g_2)\;.
\eea
Since all elements of $\Theta'^{\mu \nu}$ and $G'_{\mu \nu}$ are
finite, this is a definite noncommutativity.
\subsection{Motion along the $X^3$-direction}

According to the Lorentz transformations
\bea
&~& X'^3 = \gamma (X^3 - v\sqrt{g_0} X^0) \;,
\nonumber\\
&~& X'^0 = \gamma (X^0 - \frac{v}{\sqrt{g_0}}X^3) \;,
\eea
the noncommutativity parameter $\Theta^{\mu \nu}$ does not change,
while the open string metric $G_{\mu \nu}$ transforms to
\bea
G'_{\mu \nu} = \left( \begin{array}{ccc}
-\gamma^2 g_0 & 0 & 0 \\
0 & g_1 a & g' a \\
0 & g' a & g_2 a
\end{array} \right)\;.
\eea
From the relativistic point of view,
the motion perpendicular to the brane
does not change the lengths along the brane. Furthermore,
since the noncommutativity due to the magnetic field only has the
space-space elements, it is an expected result that 
$\Theta^{\mu \nu}$ and the space-space elements of $G_{\mu \nu}$
remain unchanged.

For special speeds of the membrane there is
$\gamma^2 = a$, or equivalently
\bea
v = \pm \frac{2\pi \alpha' b}{\sqrt{g+(2\pi \alpha'b)^2}}\;.
\eea
These speeds are not greater than the speed of light.
In these frames the open string metric reduces to 
\bea
G'_{\mu \nu} = a g_{\mu \nu}\;.
\eea
Therefore, the open string metric is the scaled closed string
metric by the scale factor $a$.
\section{Electric field noncommutativity}

Now consider an electric field along the $X^1$-direction i.e., $E' = b=0$.
The noncommutativity parameter of this system is
\bea
\Theta^{\mu \nu} = T(E) \left( \begin{array}{ccc}
0 & g_2 & -g' \\
-g_2 & 0  & 0 \\
g' & 0 & 0 
\end{array} \right)\;,
\eea
where the function $T(E)$ is
\bea
&~& T(E) = \frac{1}{g_2}\frac{E}{E_0^2- E^2}\;,
\nonumber\\
&~& E_0 \equiv \frac{1}{2\pi \alpha'}\sqrt{\frac{g_0 g}{g_2}}\;.
\eea
The electric field $E$ only admits space-time noncommutativity.
Since all elements of the noncommutativity parameter are proportional to 
the function $T(E)$, this factor shows the strength of the noncommutativity.

The effective open string coupling is
$G_s = g_s \sqrt{1- \bigg{(}\frac{E}{E_0}\bigg{)}^2}$.
According to the root factor there is
$-E_0 \leq E \leq E_0$. Therefore, unlike $\theta(b)$, the function
$T(E)$ is one to one i.e., each noncommutativity strength only
corresponds to one value of the electric field.

Near the electric field $E_0$, open
strings do not interact and the strength
of the noncommutativity is infinite.
Although at the critical electric field $E_0$ the effective
theory of the open string is singular and
ill-defined, it is possible to define a
space-time noncommutative theory by taking an appropriate scaling limit.
\subsection{Motion in the $X^1$-direction}

The transformations (7) give the noncommutativity parameter as
in the following
\bea
\Theta'^{\mu \nu} = T(E) \left( \begin{array}{ccc}
0 & g_2 & -\gamma g' \\
-g_2 & 0  & \gamma v g' \sqrt{g_0} \\
\gamma g' & -\gamma v g' \sqrt{g_0} & 0 
\end{array} \right)\;.
\eea
The open string metric also has the transformation
\bea
&~& G'_{00} = -\gamma^2 g_0 \bigg{[} 1-g_1 v^2 -\bigg{(}1-\frac{g}{g_2}v^2
\bigg{)}\frac{E^2}{E_0^2}\bigg{]}\;,
\nonumber\\
&~& G'_{01} = -\gamma^2 v \sqrt{g_0} \bigg{[} 1-g_1 -\bigg{(}1-\frac{g}{g_2}
\bigg{)}\frac{E^2}{E_0^2}\bigg{]}\;,
\nonumber\\
&~& G'_{02} = \gamma v g' \sqrt{g_0} \;,
\nonumber\\
&~& G'_{11} = \gamma^2 \bigg{[} g_1- v^2 -\bigg{(}\frac{g}{g_2}-v^2
\bigg{)}\frac{E^2}{E_0^2}\bigg{]}\;,
\nonumber\\
&~& G'_{12} = \gamma g' \;,
\nonumber\\
&~& G'_{22} = g_2\;.
\eea
For $v=0$ this metric reduces to the open string metric of
the static membrane with electric field.

Near the electric field $E=E_0$ when there is $g' \rightarrow 0$,
for any value of the speed $v$, 
all elements of the transformed metric $G'_{\mu \nu}$,
except $G'_{22}$, go to zero
\bea
G'_{\mu \nu} = 0 \;\;\;\; {\rm except}\; G'_{22}\;, {\rm for} 
\;\;g' \rightarrow 0\;\;,\;\;E \rightarrow E_0\;\;.
\eea
To avoid this singularity, we can do the following 
scaling limit. For $E \rightarrow E_0$, $g' \rightarrow 0$ and finite 
$g_2$, we should have
\bea
\gamma g' = \kappa\;,\;\;\;\;\gamma^2\bigg{(}1-\frac{E^2}{E_0^2}
\bigg{)} = \rho\;,
\eea
where $\kappa$ and $\rho$ are finite constants. These imply that the
boost velocity approaches to the speed of light, $v \rightarrow \pm 1$.
Therefore, the metric (22) takes the form
\bea
G'_{\mu \nu} = \left( \begin{array}{ccc}
-\rho g_0(1-g_1) & -\rho {\sqrt g_0}(1-g_1) & \kappa{\sqrt g_0} \\
-\rho {\sqrt g_0}(1-g_1) & -\rho (1-g_1)  & \kappa  \\
\kappa{\sqrt g_0} & \kappa  & g_2 
\end{array} \right)\;.
\eea
All elements of this metric are finite. To restore interactions of open
strings, the string coupling $g_s$ can be scaled to infinity i.e., 
$g_s \sim \gamma$. This leads to a finite $G_s$. The noncommutativity
parameter $\Theta'^{\mu\nu}$ near the critical field
$E_0$ also should be finite. Therefore,
we scale $\alpha'$ to zero as $\alpha' = \mu (1- \frac{E^2}{E_0^2})$,
where $\mu$ is finite. In other words, after scaling we have
\bea
\Theta'^{\mu \nu} = \frac{2\pi \mu}{\sqrt{g_0g_2g}}
\left( \begin{array}{ccc}
0 & g_2 & -\kappa \\
-g_2 & 0 & \kappa \sqrt{g_0} \\
\kappa & -\kappa \sqrt{g_0} & 0 
\end{array} \right)\;,
\eea
which describes a well defined noncommutativity.
\subsection{Motion in the $X^3$-direction}

From the transformations (15) we obtain 
${\Theta'}^{\mu\nu} = \gamma {\Theta}^{\mu\nu}$, where
${\Theta}^{\mu\nu}$ has been given by the equation (19).
The open string metric also becomes
\bea
G'_{\mu \nu} = \left( \begin{array}{ccc}
-\gamma^2 g_0 (1- \frac{E^2}{E_0^2}) & 0 & 0 \\
0 & g_1 (1-\frac{g}{g_1 g_2}\frac{E^2}{E_0^2})  & g'  \\
0 & g'  & g_2 
\end{array} \right)\;.
\eea
Again in the limit $E \rightarrow E_0 $ but arbitrary $v$ and $g'$, 
we should introduce a scaling limit. Let the elements of the closed string 
metric $g_0$, $g'$ and $g_2$ be finite. In the limit
$E \rightarrow E_0$ we can put $\gamma^2 (1-\frac{E^2}{E_0^2}) = \rho$,
which gives 
\bea
G'_{\mu \nu} = \left( \begin{array}{ccc}
-\rho g_0 & 0 & 0 \\
0 & \frac{g'}{g_2} & g'  \\
0 & g'  & g_2 
\end{array} \right)\;.
\eea
In this limit, the speed $v$ approaches to the speed of light such
that $\rho$ to be finite.

To have a finite noncommutativity parameter, the parameter
$\alpha'$ should go to zero like 
$\alpha' = \beta(1- \frac{E^2}{E_0^2})^{3/2}$, where $\beta$ is another 
finite constant. Therefore, we obtain 
\bea
\Theta'^{\mu \nu} = 2\pi \beta \sqrt{\frac{\rho}{g_0g_2g}}
\left( \begin{array}{ccc}
0 & g_2 & -g' \\
-g_2 & 0 & 0 \\
g' & 0 & 0 
\end{array} \right)\;.
\eea

Since $\alpha'$ goes to zero, according to the definition (20),
$E_0$ approaches to infinity. In this limit let $E$ be
proportional to $E_0$ as in the following
\bea
E = \pm \sqrt{\frac{1}{2}\bigg{(}1+\sqrt{1-4q^2 \rho}\bigg{)}}\;E_0\;,
\eea
where $q$ is an infinitesimal number from the interval
$-\frac{1}{2 \rho^{1/2}}\leq q \leq \frac{1}{2\rho^{1/2}}$.
Using the definition of $\rho$,
this equation also can be written in the form $E = q \gamma E_0$.
This implies that $q$ goes to zero such that $q\gamma \rightarrow \pm 1$.
In other words, the field $E$ approaches to infinity like $\gamma^3$,
\bea
E = \frac{q \gamma^4}{2\pi \beta \rho^{3/2}}\sqrt{\frac{g_0 g}{g_2}}\;.
\eea
For this large electric field, the noncommutativity parameters (19)
and (29) are related to each other through the equation
\bea
\Theta'^{\mu \nu}_{\rm scaled} = \frac{1}{q} \Theta^{\mu \nu}_E\;. 
\eea
Since there is $q \neq 1$, the above noncommutativities are equivalent but
cannot be equal.
\subsection{Equivalence of the electric and magnetic noncommutativities}

Consider two parallel membranes which move along the $X^1$-direction.
The first is a $D2$-brane with the speed $v$ and the magnetic field $b$.
Its corresponding closed string metric has the elements
$(-g_0 , g_1 , g_2=0 , g')$. The noncommutativity of this
brane has been given by the equation (8). The second is $D2'$-brane
with the speed $v'$, electric field $E$ and the closed string metric
elements $(-g'_0 , g_1 , g_2=0 , g')$. 
This system has the noncommutativity parameter (21).

It is possible to have the relation
\bea
\Theta'^{\mu \nu}_E = \eta \Theta'^{\mu \nu}_b\;, 
\eea
where $\eta$ is any real number. This matrix equation leads to the
conditions
\bea
&~& \frac{T(E)}{\theta(b)} = \frac{\eta}{g'} \gamma \sqrt{\frac{v^2}{g_0}
-\frac{1}{g'_0}}\;,
\nonumber\\
&~& vv' = \sqrt{\frac{g_0}{g'_0}}\;,
\eea
where $\theta(b) = \frac{(2\pi \alpha')^2b}{g'^2-(2\pi \alpha' b)^2}$
and $T(E) = -(2\pi \alpha')^2 \frac{E}{g'_0 g'^2}$ with
$-\infty < E< +\infty$. On the other hand, electric field also
does not have critical value.
The first condition compares the strengths of the noncommutativities.
The second condition imposes that
$g'_0 > g_0$ and the motions should be in the same direction.
For the non-zero magnetic field when $v \rightarrow \pm 1$ we obtain
$E \rightarrow \infty $, which is available. For $\eta = 1$ these
different systems completely have the same noncommutativity structure.
\section{$T$-duality and equivalence of noncommutativities}

In the case of toroidal
compactification, when $d$-spatial coordinates are compactified on torus
$T^d$, the $T$-duality group is $O(d,d; {\bf Z})$ \cite{5}. Assume that a
$Dp$-brane is wrapped on torus $T^p$. Under the action of a particular 
element of $O(p,p; {\bf Z})$ $T$-duality group i.e., 
\bea
T = \left( \begin{array}{cc}
{\bf 0} & {\bf 1}_{p \times p} \\
{\bf 1}_{p \times p}  & {\bf 0} 
\end{array} \right)\;,
\eea
the background fields have the transformations \cite{6}
\bea
(g+2\pi \alpha' B) \rightarrow ( {\tilde g}+ 2\pi
\alpha' {\tilde B})=(g+2\pi \alpha' B)^{-1}\;.
\eea
According to the equations (1) and (36) we obtain
\bea
&~& G^{\mu\nu} = {\tilde g}^{\mu\nu}\;,
\nonumber\\
&~& \Theta^{\mu\nu} = (2 \pi \alpha' )^2 {\tilde B}^{\mu\nu}\;.
\eea
That is, the open string metric and the noncommutativity parameter
appear as the background fields of the $T$-dual theory
of string theory. One can show that the 
effects of $T$-duality transformations on the 
open string metric $G^{\mu\nu}$ and 
on the noncommutativity parameter $\Theta^{\mu\nu}$ are as in
the following
\bea
&~& {\tilde G}^{\mu\nu} = g^{\mu\nu}\;,
\nonumber\\
&~& {\tilde \Theta}^{\mu\nu} = (2\pi \alpha')^2 B^{\mu\nu}\;.
\eea
Therefore, the background fields of string theory appear as the effective
metric and noncommutativity parameter of the 
effective theory of the $T$-dual theory \cite{7}.

Now we find the background fields and the speed of the membrane such that
noncommutativity in the moving frame be equivalent to the
noncommutativity of the $T$-dual theory in the rest frame i.e.,
\bea
\Theta'^{\mu\nu} = - \lambda {\tilde \Theta}^{\mu\nu}\;,
\eea
where $\lambda$ is a positive constant. This means that, $T$-duality
also can act as Lorentz transformations and vice-versa.
In other words, noncommutativity parameter
in the moving frame is proportional to the background field 
$B^{\mu \nu}$ in the rest frame. 
The equations (38) and (39) give the following table for various values
of $g'$, $v$ and $B_{\mu\nu}$,

\begin{table}[ht]
\vspace{0.3cm}
\begin{center}
\begin{tabular}{|c|c|c|c|c|}
\hline
$ $ & ${\rm magnetic\;and }\;v_1$&$ {\rm magnetic\;and}\;v_3$& 
${\rm electric\;and}\;v_1$& ${\rm electric\;and}\;v_3$\\
\hline
$g'$&$g'$&$g'$&$0\;\;\;\;\;\bigg{|}\;\;\;\;g'$&$g'$\\
\hline
$v$&$0$&$v$&$v\;\;\;\;\bigg{|}\;\;\;\;0$&$v$\\
\hline
$B_{\mu\nu}$&$ b=\pm \frac{1}{2\pi \alpha'}\sqrt{(\frac{1}{\lambda}-1)g}$
&$b=\pm \frac{1}{2\pi \alpha'}\sqrt{(\frac{1}{\lambda}-1)g}$&
$E=\pm E_0 \sqrt{1-\frac{1}{\lambda}}$&
$E=\pm E_0 \sqrt{1-\frac{\gamma}{\lambda}}$\\
\hline
$G_s$&$\frac{g_s}{\sqrt{\lambda}}$&$\frac{g_s}{\sqrt{\lambda}}$&
$\frac{g_s}{\sqrt{\lambda }}$&
$g_s \sqrt{\frac{\gamma}{\lambda }}$\\
\hline
\end{tabular}
\end{center}
where $v_i$ shows the membrane motion along the $X^i$-direction.
\end{table}

When the equation (39) holds, to find the corresponding
$G'_{\mu \nu}$, we should use the equations (9), (16), (22) and (27) and
the values of $g'$, $v$, $b$ and $E$ of this table. 
For example for the magnetic part of 
the table, the space-space elements of the
metric $G'_{\mu \nu}$ are proportional to $g_{\mu \nu}$ with the scale factor
$\frac{1}{\lambda}$. For the special case that has been given by the 
equation (18) we have $\lambda=\frac{1}{\gamma^2}$. According to 
the equation (38), the equation (18) can be written as 
\bea
G'_{\mu \nu} = \frac{1}{\lambda} {\tilde G}_{\mu \nu}\;.
\eea
That is, on the open string metric, 
Lorentz transformations also act as $T$-duality.

Noncommutativity parameter in the moving frame is $\Theta'^{\mu \nu}=
-(2\pi \alpha')^2 \lambda B^{\mu \nu } $.
For the magnetic case ${\Theta'}^{\mu \nu }$ is similar to the matrix (4),
that $\theta$ should be replaced with the factor
$\mp 2\pi \alpha' \sqrt{\frac{\lambda -\lambda^2}{g}}$. 
This factor is the strength of the noncommutativity, and for
$\lambda = \frac{1}{2}$ has extremum $\mp \frac{\pi \alpha'}{\sqrt{g}}$.
For the electric case $\Theta'^{\mu \nu}$ is given by the matrix
(19) in which the function $T(E)$ should be changed with the factor
$\pm \frac{1}{g_2 E_0} \sqrt{\lambda^2 - \eta \lambda}$. For the motion
along the $X^1$ and $X^3$ directions $\eta$ is 1 and $\gamma$, respectively.

Note that the equation $\Theta^{\mu\nu} = 
- \lambda ({\tilde \Theta})'^{\mu\nu}$, for the 
membrane with magnetic field, produces the results of the equation (39).
The matrix $({\tilde \Theta})'^{\mu\nu}$ is noncommutativity parameter
of the $T$-dual theory in the moving frame.
\section{Conclusions}
For the moving membranes with electric or magnetic background
fields, we studied the effective variables of open string.
For the magnetic membrane we observed the followings.
There are two values of the magnetic
field that produce the same noncommutativity on the membrane.
By choosing an appropriate magnetic field, the open string metric
becomes independent of the speed of the membrane.
When this magnetic field goes to zero and near the speed of light
we obtained a well defined noncommutativity. For the special
speeds perpendicular to the membrane, the open string metric is
proportional to the closed string metric.

For the pure electric field on the membrane, the strength of 
the noncommutativity in terms of the electric field is
one to one. When the electric field 
approaches to its critical value, we obtained some definite
noncommutative theories from decoupling limits.
We found the conditions that the noncommutativity structures of the
moving electric and magnetic membranes to be equivalent.
Similarly the equivalence of the electric noncommutativities in the
rest frame and in the moving frame (with the scaling limit form)
was obtained.

The background fields of string theory ($T$-dual of string 
theory), are effective metric and noncommutativity parameter of 
the effective $T$-dual theory (the effective theory of string theory). 
Therefore, we observed that for the special background fields and speeds 
of the membrane, the noncommutativity of the $T$-dual theory in the
rest frame appears like the noncommutativity of the original
theory in the moving frame. The open string metric
in the moving frame also is equivalent to that one
of the $T$-dual theory in the rest frame.

\end{document}